# Dual-Energy Cone-Beam CT Using Two Orthogonal Projection Views: A Phantom Study

Junbo Peng[1], Tonghe Wang[2], Shaoyan Pan[1], and Xiaofeng Yang[1,*]

[1]Department of Radiation Oncology and Winship Cancer Institute, Emory University, Atlanta, GA 30322

[2]Department of Medical Physics, Memorial Sloan Kettering Cancer Center, New York, NY 10065

Email: xiaofeng.yang@emory.edu

## Abstract

This study proposes a novel imaging and reconstruction framework for dual-energy cone-beam CT (DECBCT) using only two orthogonal X-ray projections at different energy levels (2V-DECBCT). The goal is to enable fast and low-dose DE volumetric imaging with high spectral fidelity and structural accuracy, suitable for DECBCT-guided radiation therapy. We introduce a framework for 2V-DECBCT based on physics-informed dual-domain diffusion models. A cycle-domain training strategy is employed to enforce consistency between projection and volume reconstructions through a differentiable physics-informed module. Furthermore, a spectral-consistency loss is introduced to preserve inter-energy contrast during the generative process. The model is trained and evaluated using 4D XCAT phantom data under realistic anatomical motion. The method produces high-fidelity DECBCT volumes from only two views, accurately preserving anatomical boundaries and suppressing artifacts. Subtraction maps computed from the reconstructed energy volumes show strong visual and numerical agreement with ground truth. This work presents the first diffusion model-based framework for 2V-DECBCT reconstruction, demonstrating accurate structural and spectral recovery from extremely sparse inputs.

# 1. Introduction

Cone-beam CT (CBCT) has been a well-established imaging modality in current image-guided radiation therapy workflow. It is used as a standard practice for patient setup verification.[1] However, conventional single-energy CBCT has limitations in image artifacts and poor soft-tissue contrast stemming from scattered radiation and beam-hardening.[2] The non-uniformity and inaccuracies in Hounsfield Units (HU) not only degrade soft-tissue visibility but also pose challenges for quantitative usage in adaptive radiation therapy (ART).

Dual-energy CBCT (DECBCT) has been proposed to overcome the limitations of single-energy imaging and extract more quantitative information from scans. Observing the same principle as dual-energy CT (DECT), DECBCT scans objects with two different X-ray energy spectra in order to provide material-specific contrast, thereby enhancing guidance capability and enabling adaptive treatment strategies.[3] Preliminary studies have demonstrated that DECBCT can significantly enhance tissue contrast, improve tumor visualization, and enable precise material decomposition—capabilities that are essential for accurate online segmentation and dose calculation.[4] For example, in proton therapy, where dose distribution is highly sensitive to tissue composition, having DECBCT-derived electron density maps each day could allow clinicians to adjust beam energies or angles to account for anatomical changes, thereby optimizing the delivered dose to the tumor while minimizing range uncertainties.[5] Hardware-based approaches to DECBCT have typically involved prototyping dual-layer flat-panel detectors (FPDs), but these systems require sophisticated hardware and are limited by insufficient spectral separation between the detector layers.[6,7] An alternative approach using fast-kVp switching has been implemented in LINAC-mounted CBCT systems, offering improved energy separation and demonstrating promising results for markerless tumor tracking.[8,9] This technique does not require hardware upgrades, making it more feasible for clinical integration. Recently, deep learning has further enabled single-scan DE imaging with only one kVp switch, simplifying implementation and reducing the technical burden.[10] However, a shared limitation among these methods is their reliance on full-trajectory scanning—typically requiring about one minute to acquire X-ray projections from hundreds of angles, similar to conventional single-energy CBCT. This prolonged acquisition time increases susceptibility to patient motion artifacts, which can degrade image quality and compromise the accuracy of quantitative imaging tasks. To overcome these challenges, innovative strategies are urgently needed to enable fast, low-dose DECBCT solutions that maintain image quality while minimizing motion sensitivity, ideally without the need for major hardware modifications.

In this study, we propose to use AI to generate volumetric DECBCT images from orthogonal X-ray projections acquired at two different energy levels (2V-DECBCT). By acquiring only two orthogonal X-ray projections to generate DECBCT volume, we expect it can avoid the long scanning time in current approaches and therefore reduce patient motion artifacts. Moreover, the technique is particularly well-suited for advanced radiotherapy platforms such as ProBeam and CyberKnife, where precise localization and rapid online adaptation are critical.[11] Successful implementation of this technology will represent a significant advancement in DECBCT-based online ART and broaden its clinical applicability across multiple treatment systems.

Generating 3D volumetric images from 2D projections using the patient-specific model has been investigated under different conditions. For example, the authors utilized V-shaped convolutional neural network (CNN) to reconstruct CBCT images from a single-view projection.[12] In ref,[13] the authors proposed a novel perceptual supervision-based generative adversarial network (GAN) to derive instantaneous volumetric images from a single projection. In ref,[14] the authors proposed the geometry-integrated cycle-domain diffusion models to reconstruct volumetric images from a single-view projection for real-time image guidance in radiation therapy. In ref,[15] the authors designed a CNN with feature extractors and residual blocks to reconstruct 3D images from orthogonal projections for the ProBeam system. However, these techniques have not been used for DECBCT imaging with ultra-sparse data acquisition, where spectral contrast information is critical in the material quantification tasks. In this study, we propose a physics-integrated dual-domain diffusion models (DM) with a spectral contrast-preserving strategy to generate DECBCT images from two orthogonal DE projections. This proof-of-concept study is expected to pave the way for fast and low-dose spectra volumetric imaging to enhance the clinical feasibility and safety of high-frequency imaging in IGRT and ART workflows.

## 2. Methodology

### 2.1 Overview of Framework

The overall framework of our 2V-DECBCT is shown in Figure 1, which includes: i) projection-domain DM, which synthesizes full-view DE X-ray projections from the orthogonal-view DE projections; ii) image-domain DM, which reconstructs the DECBCT from noisy samples, conditioned on priors; and iii) physics-integrated module, which implements differentiable forward projection and reconstruction operators to transfer information between the two domains. The dual-domain diffusion models are synergetically trained to minimize both reconstruction and data-consistency errors under an interactive-domain supervisory scheme.

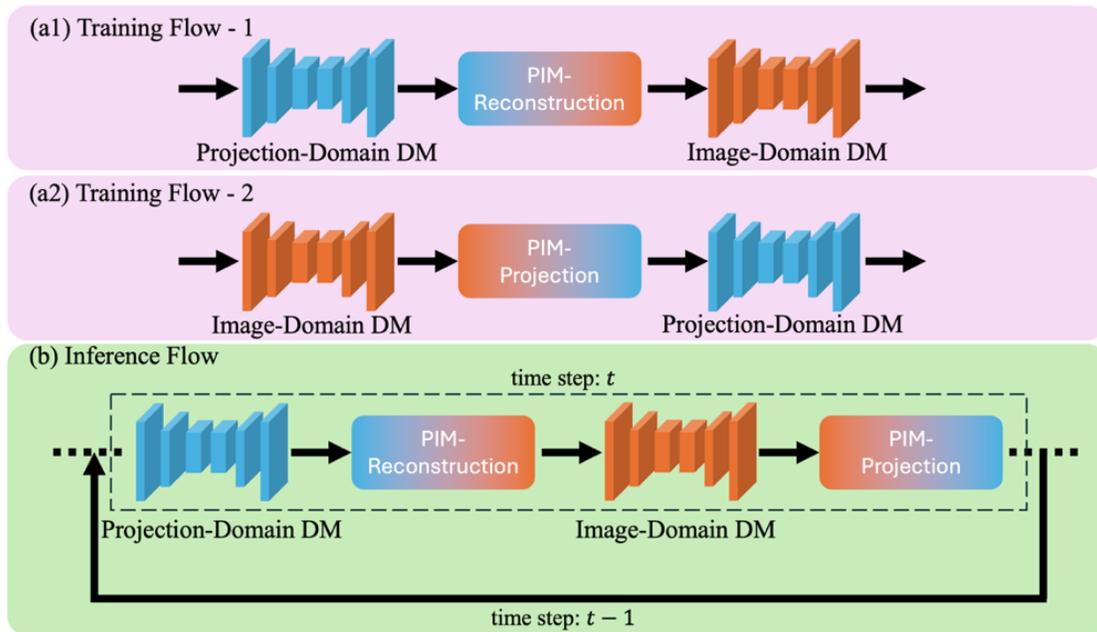

**Figure 1.** Overview of the 2V-DECBCT framework.

### 2.2 Improved Denoising Diffusion Probabilistic Model (iDDPM)

To enhance the stability and flexibility of the reverse sampling process, we adopt the Improved Denoising Diffusion Probabilistic Model (iDDPM) framework in both image-domain and projection-domain DMs.[16] In contrast to the original DDPM formulation,[17] iDDPM introduces a learnable variance interpolation mechanism that allows the model to predict both the mean and variance of the reverse distribution at each diffusion timestep. This improvement has been shown to yield higher sampling quality and reduced artifacts, particularly under limited conditional supervision.

Specifically, given a noisy sample $X_t$ and conditioning input $Z$ (e.g., the coarse DECBCT or orthogonal-view DE projections), the model predicts the clean latent $\hat{X}_0 = \theta_\mu(X_t, t, Z)$, which is used to compute the mean of the reverse transition:

$$\mu_\theta(X_t, t, Z) = \frac{1}{\sqrt{\alpha_t}}\left(X_t - \frac{1-\alpha_t}{\sqrt{1-\bar{\alpha}_t}}\hat{X}_0\right) \tag{1}$$

In the iDDPM, the reverse variance $\Sigma_\theta$ is also learned through interpolation between fixed and trainable components. This allows the network to express uncertainty adaptively at each timestep, improving both reconstruction quality and sampling stability.

## 2.3 Differentiable Physics-Informed Module (PIM)

The physics-informed module (PIM) bridges the models and data in projection and image domains, which simulates the DECBCT reconstruction and projection operators in a differentiable manner and enables the gradient propagation between the cyclic domains during the model training. The forward and backward pairs in the PIM model finite-size voxel and detector pixel, which is similar to the separable footprint method. The matched forward and backward operators ensure convergence and provide accurate, smooth results.

## 2.4 Model Training

The proposed model is trained using a supervised learning strategy under a probabilistic denoising diffusion framework. Given paired input-output data samples, the model learns to reverse a predefined forward diffusion process by estimating the conditional posterior distribution at each diffusion timestep. To guide the learning process and enforce anatomical and geometric consistency, we design a composite loss function that integrates noise prediction, uncertainty modeling, and domain-specific constraints.

The training procedure alternates between CBCT and projection denoising paths, incorporating strict data consistency via a PIM. At each training iteration, a clean target $X_0$ is corrupted to a noisy observation $X_t$ via a fixed forward process. The model is trained to estimate the denoising mean $\hat{X}_0$ and adaptively interpolates the reverse variance $\Sigma_t$ using time step- and condition-dependent features. The predicted mean is used to construct the reverse transition distribution, while the variance interpolation is regularized via a variational divergence loss.

To ensure consistency across domains, we incorporate both direct reconstruction and measurement-consistency supervision. For example, in volumetric reconstruction tasks, the model is penalized not only based on the fidelity of the denoised DECBCT image but also on its ability to generate consistent projection data via differentiable forward modeling. Conversely, projection outputs are also constrained by their backprojected counterparts through physics-informed supervision.

Formally, the total training objective is defined as:

$$\mathcal{L}_{\text{total}} = \frac{1}{2}\left(\mathcal{L}_{\text{img}} + \mathcal{L}_{\text{proj}}\right) \tag{2}$$

## 2.5 Inference Procedure

At inference time, the proposed framework reconstructs a volumetric CBCT image from a single-view projection by progressively denoising a sequence of variables through a fixed number of diffusion steps. At each timestep, the denoised DECBCT is forward-projected via the differentiable PIM to yield a synthetic projection stack, which serves as a strict condition for projection-domain DM. Simultaneously, the updated projection is back-projected to generate a volumetric prior, which is then fed into image-domain DM. This reciprocal conditioning mechanism continues across all remaining timesteps, enabling the two networks to reinforce anatomical and geometric consistency mutually. The final output is a high-fidelity DECBCT volume that leverages both learned structural priors and measurement-consistent projections. This iterative strategy improves the robustness of reconstruction under extreme view sparsity and ensures that the output is consistent with both imaging physics and anatomical plausibility.

## 3. Experiments

3.1 Dataset

In this work, we employed the extended cardiac–torso (XCAT) phantoms, which are highly realistic anthropomorphic digital phantoms widely used for simulation-based validation in medical imaging.[18] Four digital phantoms were used in this study. The 4D XCAT model provides anatomically accurate representations of human anatomy and incorporates physiologically plausible respiratory and cardiac motion, enabling the generation of time-resolved volumetric CT data with known ground truth. For each phantom, eight phases were used for patient-specific model training, one phase was used for model validation, and one phase was used for testing.

3.2 4D-CT Simulation

4D-DECBCT data is required for model training and validation in the proposed method. However, only the 4D-CT dataset can be acquired in clinical practice. To address this issue, we generate the 4D-DECBCT dataset from the 4D-CT dataset following two steps. In the first step, we segment the 4D-CT images into bone, muscle, and adipose according to the HU values. Thus, we obtain three material-specific maps for each phase. Next, we simulate the two full CBCT scans under high- and low-kVps to acquire the DE-CBCT images of each phase.

Using the generated XCAT digital phantom in each phase, the 4D-CT projection was simulated using

$$\vec{I} = \int S(E) e^{-\sum_i \{\mu_i(E) \int \vec{x}_i d\vec{l}\}} dE \qquad (3)$$

where where $\vec{I}$ is the detected photon number after transmission of the phantom, $S(E)$ is the spectrum of the incident X-ray beams, $\mu_i(E)$ is the linear attenuation coefficient (LAC) of the $i$-th basis material at energy $E$, $\vec{x}_i$ is the volumetric map of the $i$-th material, and $\vec{l}$ is the transmission path of each X-ray beam. In this way, the spectral effect (e.g., beam-hardening effect) can be incorporated into the generated projections. In this study, a quantized formula was used for the simulation as

$$\vec{I} = \sum_E S(E) e^{-\sum_i \{\mu_i(E) F \vec{x}_i\}} \qquad (4)$$

where $S(E)$ is the discrete X-ray spectrum and $F$ indicates the forward projection operator corresponding to the scanning geometry.

The X-ray spectrum was generated using Spektr toolbox,[19] which involved 0.5-mm aluminum (Al) for the equivalent inherent filtration and 0.3-mm copper (Cu) as the additional filtration to block low-energy photons. The tube voltage was set to 100 kVp to align with the Siemens SOMATOM Definition AS scanner in our institute. In addition, Poisson noise is added to the detected photon number $\vec{I}$ with an incident photon number ($I_0$) of $1 \times 10^5$ at each detector pixel. The acquired line integrals (LI) can be calculated using

$$\vec{LI} = \frac{\vec{I}}{I_0} = \frac{\vec{I}}{\Sigma_E S(E)} \tag{5}$$

Then, the 4D-CT images were reconstructed from the acquired line integrals.

### 3.3 4D-DECBCT Simulation

With the generated 4D-CT images in the previous step, we first segment the images into bone, muscle, and adipose maps according to the HU values. Then, we simulated DECBCT projections on the 4D material-specific maps with the DE X-ray spectra at 70 kVp and 140 kVp using a similar process in the previous section.

### 3.4 Implementation Details

Both image- and projection-domain DMs were equipped with timestep embeddings and conditional feature modulation via concatenated inputs. The input to each model consisted of a noisy sample concatenated with its respective conditioning tensor along the channel dimension. The PIM was implemented using the LEAP toolbox.[20] All models were implemented in PyTorch and trained using the Adam optimizer. Training was performed using paired DECBCT volumes and full-view projections, with data augmentation applied in both domains. The batch size was set to 2, and training was conducted for 200k iterations using mixed precision on NVIDIA A100 GPUs.

### 3.5 Evaluations

To validate the effectiveness of our approach, we compared the reconstructed DECBCT volumes against ground-truth references generated from full-view dual-energy data. Results were evaluated separately for low-energy and high-energy channels. Mean absolute error (MAE), normalized correlation coefficient (NCC), and structural similarity index measurement (SSIM) were used for quantitative evaluation across both energy levels.

## 4. Results

Representative slices from the reconstructed DE-CBCT volumes are shown in Fig. 2. The proposed method preserves fine structural details and suppresses streaking artifacts commonly observed in sparse-view FDK reconstructions. Notably, soft-tissue boundaries are clearly delineated in both energy levels. The MAE of all 400 slices of the four phantoms was $17.96 \pm 2.98$ HU for the high-energy images and $24.79 \pm 4.73$ HU for the low-energy images. The calculated NCC was $0.98 \pm 0.02$ for the high-energy images and $0.97 \pm 0.02$ for the low-energy images. The SSIM was $0.99 \pm 0.01$ and $0.98 \pm 0.01$ for the high- and low-energy levels.

In addition to individual energy reconstructions, we visualized the energy subtraction maps computed by subtracting the high-energy CBCT from the low-energy CBCT. This differential map enhances material contrast and is particularly effective in highlighting tissue compositions with distinct attenuation responses across energies. As shown in Fig. 4, the subtraction maps generated by the proposed method exhibit high spatial agreement with the ground-truth subtraction maps obtained from full-view reconstructions. The differences between our predicted and reference subtraction maps are visually minimal, indicating that both anatomical alignment and spectral fidelity are well preserved. This consistency suggests that our method not only reconstructs each energy volume accurately but also maintains reliable inter-energy contrast behavior.

To demonstrate the contrast enhancement in the energy subtraction maps, we calculated the relative contrast of a low-contrast ROI in the phantoms. The relative contrasts were 0.15/0.20 in the ground-truth high-/low-energy images and 0.14/0.20 in the generated high-/low-energy images. The values were improved to 0.55 in the ground-truth subtraction maps and 0.49 in our results. These results suggest that the method not only restores structural accuracy but also recovers clinically meaningful spectral contrast.

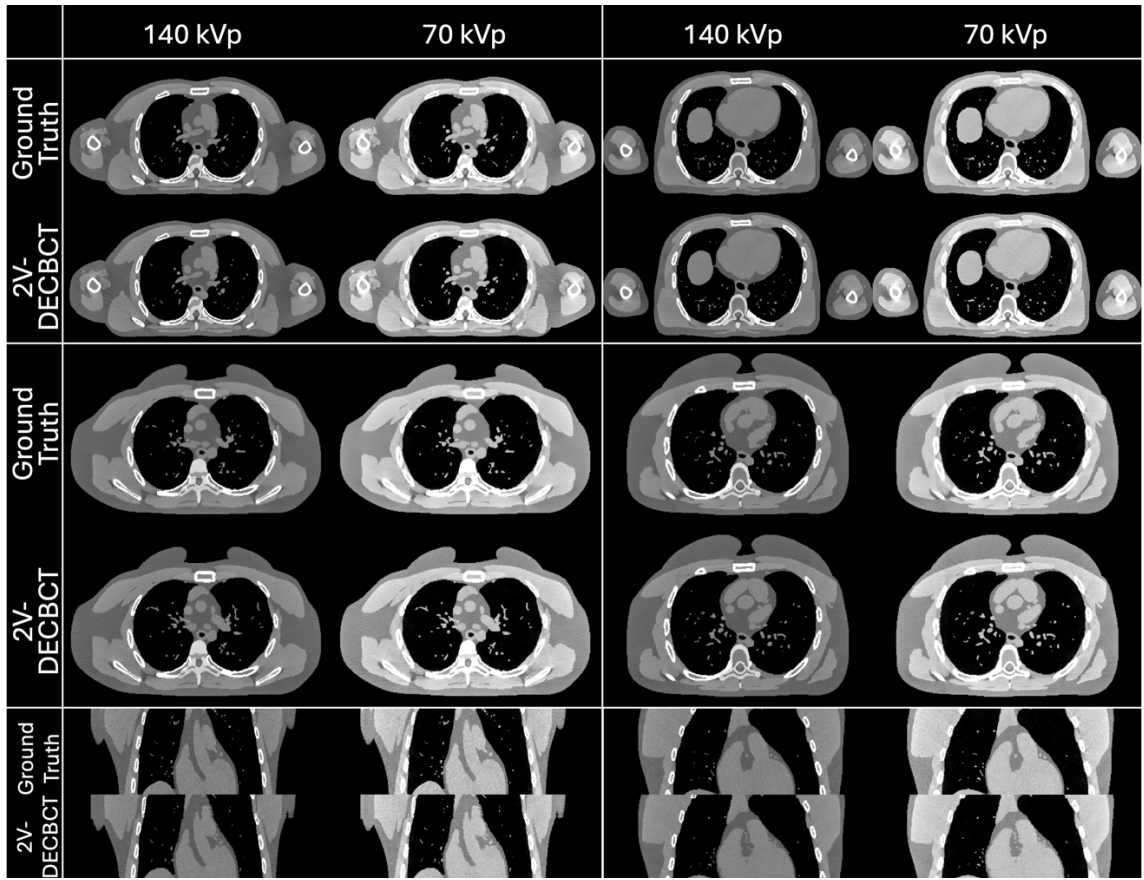

**Figure 2.** Ground truth and reconstructed 2V-DECBCT images.

## 5. Discussions

This study proposes a novel framework for reconstructing DECBCT volumes using only two orthogonal X-ray projections, 2V-DECBCT, offering an extremely sparse acquisition solution suitable for image-guided procedures with stringent dose or time constraints. To the best of our knowledge, this is the first work that formally defines the two-view DECBCT problem and provides an end-to-end learning-based solution under this highly ill-posed setting. The proposed method introduces a pair of diffusion models that operate in parallel across both the projection and image domains, corresponding to each energy channel. Through a cycle-domain training mechanism, physics consistency between projections and reconstructions is enforced at every diffusion step. In addition, we incorporate a spectral-consistency constraint into the training process, which encourages the two energy-specific networks to preserve the expected differential contrast across energy channels. This cross-energy supervision plays a critical role in maintaining spectral fidelity and producing diagnostically meaningful subtraction maps.

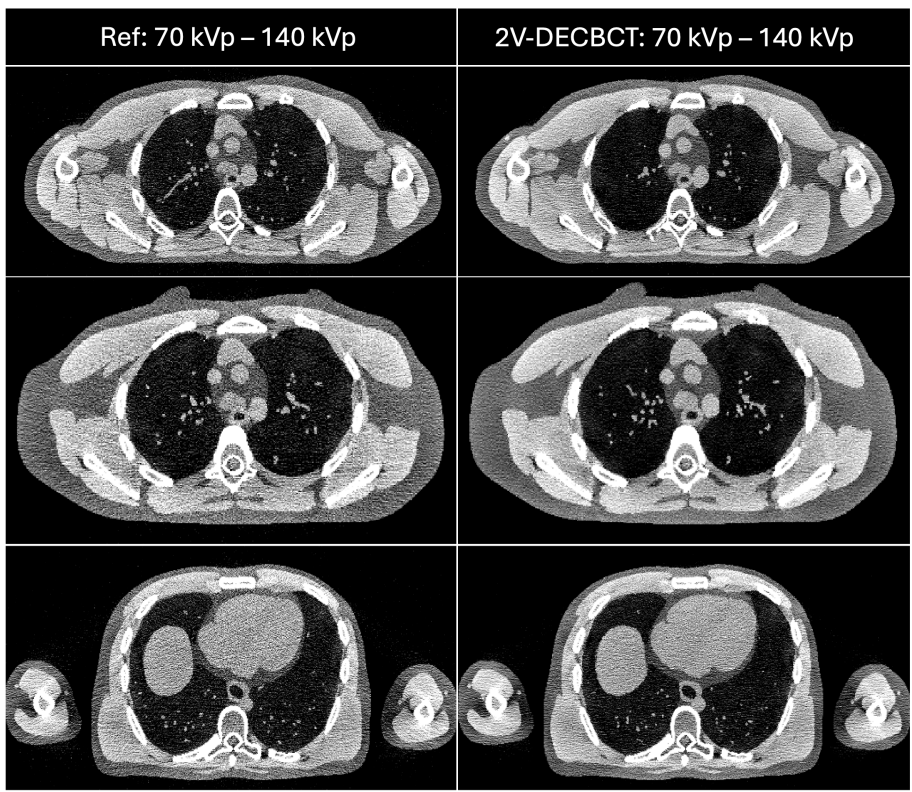

**Figure 3.** Energy subtraction maps using the ground truth and 2V-DECBCT images.

While our results demonstrate the technical feasibility and potential of 2V-DECBCT reconstruction, several limitations remain. First, the current model does not account for the presence of scattered contamination, which is a major source of image degradation in CBCT, particularly under wide-field, large-patient geometries.[21] In future studies, we plan to incorporate physically realistic scatter effects by means of Monte Carlo simulations and perform artifact reduction using image-domain methods.[22] These simulations will enable us to evaluate model robustness under realistic system physics.

Second, all experiments in this study were conducted on the 4D XCAT phantom, which offers anatomically realistic yet synthetic representations of human anatomy and respiratory motion. While this phantom enables controlled studies with known ground truth, it lacks the structural variability and pathological complexity encountered in real-world clinical data. In future work, we will extend our evaluation to include simulated patient datasets derived from CT scans of real patients, as well as physical phantom studies acquired on real scanners. These additional datasets will enable us to assess generalizability, identify failure modes under unseen anatomies, and benchmark clinical applicability in realistic scenarios.

Third, the current implementation is limited to an acquisition geometry using two orthogonal projections at 0° and 90°. In the follow-up studies, we will explore alternative dual-view configurations (e.g., 45° and 135°) to align with different machines (e.g., ProBeam and CyberKnife). This direction is essential to broaden the applicability of our method and to evaluate angular sensitivity under diverse mechanical and anatomical constraints. It will also inform the design of optimal two-view configurations that balance image quality, contrast preservation, and acquisition feasibility for different clinical applications.

## 6. Conclusions

In summary, this study presents a first-of-its-kind solution to DECBCT reconstruction from only two projection views using cross-energy, cross-domain generative modeling. This proof-of-concept work lays a solid foundation for further development of high-quality, low-exposure DECBCT imaging suitable for different clinical scenarios.

## Acknowledgment

This research is supported in part by the National Institutes of Health under Award Numbers R01CA272991, R01EB032680, R01DE033512, P30CA008748, and U54CA274513.